\documentclass[conference]{IEEEtran}
\IEEEoverridecommandlockouts
\usepackage{cite}
\usepackage{amsmath,amssymb,amsfonts}

\usepackage[hidelinks]{hyperref}
\usepackage{cleveref}
\usepackage{algorithmic}
\usepackage{xspace}
\usepackage{graphicx}
\usepackage{caption}
\usepackage{subcaption}
\usepackage{enumitem}
\setlist[itemize]{noitemsep, topsep=-8pt}
\usepackage{textcomp}
\usepackage{xcolor}
\usepackage[a4paper, total={184mm,239mm}]{geometry}
\def\BibTeX{{\rm B\kern-.05em{\sc i\kern-.025em b}\kern-.08em
    T\kern-.1667em\lower.7ex\hbox{E}\kern-.125emX}}

\newcommand{\ournameNoSpace}{\emph{ChatFuzz}} 
\newcommand{\ourname}{\ournameNoSpace\xspace}

\begin{document}

\title{\Large Beyond Random Inputs: A Novel ML-Based Hardware Fuzzing}

\author{\IEEEauthorblockN{Mohamadreza Rostami$^\text{\textdaggerdbl}$\IEEEauthorrefmark{1}$\thanks{\textdaggerdbl These authors contributed equally to this work.}$,  Marco Chilese$^\text{\textdaggerdbl}$\IEEEauthorrefmark{1}, Shaza Zeitouni\IEEEauthorrefmark{1}
    \\Rahul Kande\IEEEauthorrefmark{2}, Jeyavijayan Rajendran\IEEEauthorrefmark{2}, Ahmad-Reza Sadeghi\IEEEauthorrefmark{1}}
    \vspace{1em}
\IEEEauthorblockA{\IEEEauthorrefmark{1}Technical University of Darmstadt, \IEEEauthorrefmark{2}Texas A\&M University}}

\maketitle

\begin{abstract}
Modern computing systems heavily rely on hardware as the root of trust. However, their increasing complexity has given rise to security-critical vulnerabilities that cross-layer attacks can exploit. Traditional hardware vulnerability detection methods, such as random regression and formal verification, have limitations. Random regression, while scalable, is slow in exploring hardware, and formal verification techniques are often concerned with manual effort and state explosions.

Hardware fuzzing has emerged as an effective approach to exploring and detecting security vulnerabilities in large-scale designs like modern processors. They outperform traditional methods regarding coverage, scalability, and efficiency.
However, state-of-the-art fuzzers struggle to achieve comprehensive coverage of intricate hardware designs within a practical timeframe, often falling short of a 70\% coverage threshold.  
To address this challenge, we propose a novel ML-based hardware fuzzer, \ourname. Our approach leverages large language models (LLMs) to understand processor language and generate  data/control flow entangled yet random machine code sequences.
Reinforcement learning (RL) is integrated to guide the input generation process by rewarding the inputs using code coverage metrics.

Utilizing the open-source RISC-V-based RocketCore and BOOM cores as our testbed, \ourname achieves 75\% condition coverage in RocketCore in just 52 minutes. This contrasts with state-of-the-art fuzzers, which demand a 30-hour timeframe for comparable condition coverage. Notably, our fuzzer can reach a 79.14\% condition coverage rate in RocketCore by conducting approximately 199k test cases.

In the case of BOOM, \ourname accomplishes a remarkable 97.02\% condition coverage in 49 minutes. Our analysis identified all detected bugs by TheHuzz, including two new bugs in the RocketCore and discrepancies from the RISC-V ISA Simulator.
\end{abstract}

\section{Introduction}
Traditional hardware verification techniques are crucial for ensuring the reliability and correctness of a hardware design, the design under test (DUT), before fabrication. Among these techniques, random regression and formal verification methods are commonly employed. 
Despite its capacity to accommodate extensive hardware designs, random regression presents a notable efficiency problem as it tends to slow down when exploring the intricacies of a hardware design. Consequently, it encounters difficulties uncovering vulnerabilities within hard-to-reach critical components \cite{hardfails}. 
On the other hand, formal verification, which aims to ascertain whether a DUT complies with specified/predefined properties \cite{introformal}, is often regarded as an efficient approach for verifying the correctness of hard-to-reach hardware components. However, formal techniques rely heavily on manual effort from domain experts to define the required properties, which can be error-prone and time-consuming. Furthermore, formal verification frequently results in state explosion, rendering it impractical to verify the entire DUT comprehensively \cite{clarke2001progress}.
Hardware fuzzing has emerged as a promising approach for not only broadening the exploration of design space but also for revealing security vulnerabilities within intricate designs, including complex processors \cite{difuzzrtl,thehuzz,hypfuzz,morfuzz}. To bolster their effectiveness, hardware fuzzers harness coverage data, such as branch conditions, statements, and multiplexers' control registers or signals, for generating test cases and probing diverse hardware behaviors \cite{rfuzz,difuzzrtl,thehuzz,hw_like_sw}. When compared to traditional hardware verification techniques, hardware fuzzers have demonstrated broader coverage, enhanced scalability, and efficiency in identifying real-world vulnerabilities that have been associated with privilege escalation and arbitrary code execution attacks \cite{difuzzrtl,thehuzz,hypfuzz}. Nonetheless, state-of-the-art fuzzers struggle to achieve comprehensive coverage of intricate hardware designs within a practical timeframe, often falling short of a 
70\% coverage threshold in complex hardware such as a RISC-V RocketCore processor~\cite{rocketcore}.

\noindent
\textbf{Our Contributions.}  In this paper, we introduce \ourname, the first processor fuzzer that leverages machine learning for input generation and improvement with the help of coverage metrics, addressing a critical challenge in the field of processor fuzzing, namely, generating interdependent data/control flow entangled yet random instructions.

\textbf{Three-Step ML-Based Input Generation.} We present a three-step training process, including unsupervised learning to understand machine language structures, reinforcement learning with a disassembler for valid instruction generation, and further reinforcement learning using RTL simulation as a reward agent to improve the coverage.
    
\textbf{Significant Speed Enhancement.} \ourname demonstrably expedites enhancing condition coverage, attaining a coverage level of 74.96\% within less than one hour. In contrast, the current leading hardware fuzzer, TheHuzz \cite{thehuzz}, requires a much longer period of roughly 30 hours to achieve the same coverage, i.e., 34.6$\times$ faster. In the case of BOOM, \ourname accomplishes a remarkable 97.02\% condition coverage in 49 minutes. It is worth noting that TheHuzz exhibits greater efficiency compared to random regression techniques and is approximately 3.33$\times$ swifter than DifuzzRTL \cite{difuzzrtl}.

\textbf{Findings.} During fuzzing, \ourname detects approximately 6K mismatches and identifies more than 100 unique mismatches after automated analysis. These findings include all bugs that were detected by TheHuzz \cite{thehuzz} and two new bugs, namely the cache coherency management issue (CWE-1202) and the execution tracing (CWE-440).
Moreover, \ourname exposes deviations in the behavior of the RocketCore compared to the specifications in the RISC-V ISA. This showcases \ourname's efficiency in delving into the processor search space, thoroughly investigating even the most detailed corner cases specified in the RISC-V ISA specification.

% =======================================
% =======================================
% =======================================

\section{Background \& Related Work}

\subsection{Fuzzing}
Fuzzing provisions a large number of inputs to the program under test to uncover faults, bugs, or vulnerabilities that traditional testing methods may miss \cite{afl++}. The fuzzer may generate random, malformed, or unusual inputs to test how the program handles them. The initial set of test inputs, also known as \textit{seeds}, can be automatically generated or manually crafted by verification engineers. During each fuzzing round, the fuzzer manipulates the best test inputs from the preceding round using mutation operations like bit/byte flipping, swapping, deleting, or cloning to generate new inputs.
In recent years, fuzzing has gained significant attention from the hardware security community due to its numerous advantages over existing verification methods. In particular, fuzzing is highly automatable, cost-effective, scalable to real-world applications, and comprehensively covers the tested application. These factors have contributed to its growing popularity and adoption among researchers and practitioners in the field of software as well as hardware security \cite{rfuzz,difuzzrtl,directfuzz,thehuzz}.

\subsubsection{Processor Fuzzers} Traditional processor fuzzers such as DifuzzRTL~\cite{difuzzrtl} and TheHuzz~\cite{thehuzz} use code coverage and control register coverage as feedback to guide the mutation process. %detect vulnerabilities. 
These fuzzers generate seeds through random generation of instructions and mutate the instructions in the current input to generate new inputs. 
Recent research also led to hybrid hardware fuzzers such as HyPFuzz~\cite{hypfuzz} and PSOFuzz~\cite{psofuzz} that combine the capabilities of fuzzers with formal tools and optimization algorithms to improve the coverage achieved. However, these hybrid fuzzers also use the seed generation and mutation engines inherited from traditional processor fuzzers such as TheHuzz~\cite{thehuzz}.
While the seed generator and mutation engine in these fuzzers can identify valid instructions from the ISA, they do not have well-defined feedback to determine a meaningful sequence of instructions that will lead to deep design regions.

\subsection{Machine Learning}
\subsubsection{Reinforcement Learning (RL)}
RL is a branch of machine learning that studies how agents can learn from their actions and environment feedback to achieve a goal. RL differs from other forms of machine learning, such as supervised and unsupervised learning, in that the agent does not have access to labeled data or explicit rules but must discover the optimal behavior through trial and error.
The agent's objective is to find a policy, a function that maps each state to an action that maximizes the expected cumulative reward over time. This is achieved using various algorithms, such as policy-based or actor-critic methods.
Proximal Policy Optimization (PPO) is a family of model-free RL algorithms. 
PPO updates policy parameters for higher expected rewards based on policy gradient methods. Unlike traditional policy gradient methods, PPO employs a clipped surrogate objective function to control policy updates and prevent large deviations from the previous policy, ensuring stability and efficiency. PPO algorithms have been successfully applied to various domains, e.g., natural language generation.

\subsubsection{Large Language Models (LLMs)}
LLMs are large ML models for processing and generating natural language text. They leverage neural networks (NN), often using the transformer architecture, to learn from sequential data and capture long dependencies. LLMs can contain billions of parameters (weights), dictating how the model handles input and generates output.
LLMs are trained using different learning paradigms, from self-supervised to reinforcement learning, which means that they do not require labeled data or explicit rules but learn from the patterns and structures inherent in the text corpus. LLMs can perform various natural language processing (NLP) tasks, such as recognition, summarization, translation, prediction, and generation. LLMs are general-purpose models that can adapt to different domains and applications with minimal fine-tuning or prompt engineering. In a concurrent work, LLMs have been utilized in software fuzzing \cite{fuzz4all}. The proposed method creates test cases, particularly for fuzzing compilers, by training a large language model on a task that relies on human-defined prompts to generate and modify test cases. In contrast, our approach does not rely on human interaction during training and is additionally steered by coverage metrics.

% =======================================
% =======================================
% =======================================

\begin{figure*}
     \centering
     \begin{subfigure}[b]{0.4\linewidth}
         \centering
         \includegraphics[width=\textwidth]{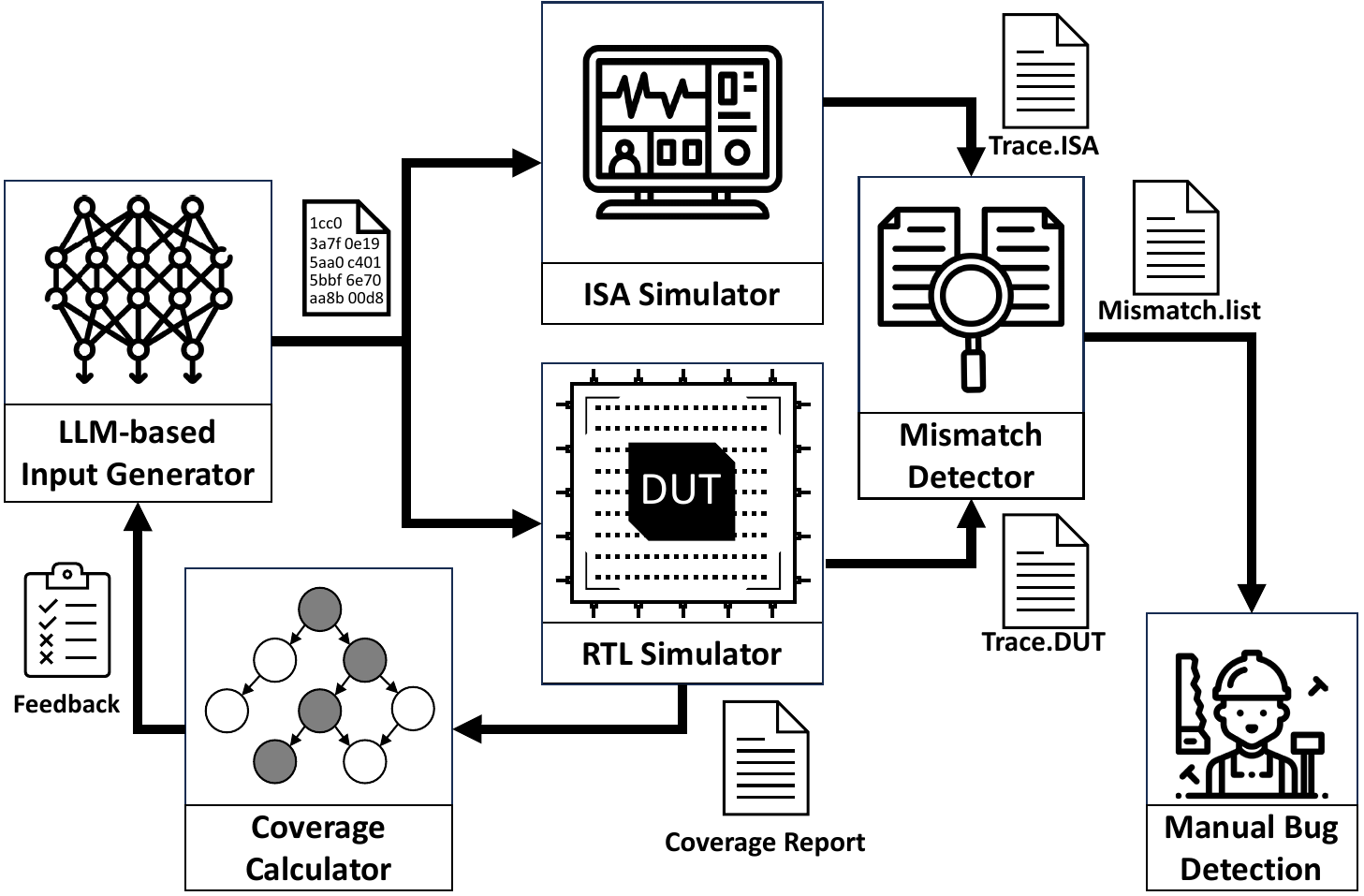}
         \caption{Overview of \ourname}
         \label{fig:overview}
     \end{subfigure}
     \hfill
     \begin{subfigure}[b]{0.55\linewidth}
         \centering
         \includegraphics[width=\textwidth]{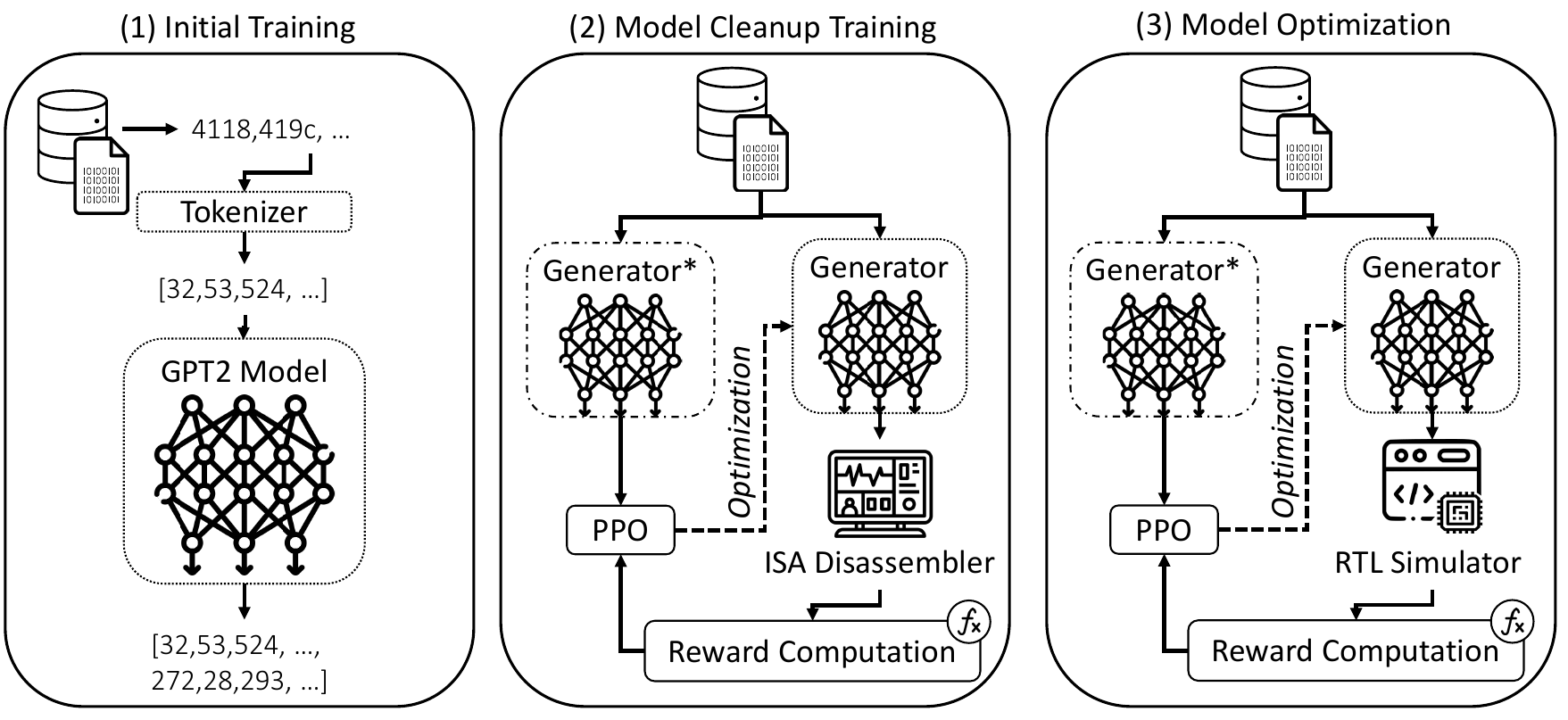}
         \caption{\ourname's training steps}
         \label{fig:system}
     \end{subfigure} 
     \subfloat{\hspace{.5\linewidth}}
     \caption{\ourname's final model results from three consequent training steps: (1) Unsupervised training based on the GPT2 model to learn the inner structure of the machine language; (2) Utilizing a disassembler as a scoring agent during PPO-based RL training, the initial model is refined by cleaning up the learned language and removing bad combinations of instructions; (3) Improving the coverage with a PPO-based RL process where the refined generator is trained through a reward function based on coverage information attained through RTL simulation.}
\end{figure*}

\section{\ourname}

Utilizing recent advancements in LLMs, we propose \ourname as an innovative approach for enhancing hardware security. \ourname involves training LLMs using machine language (specifically, machine codes) and employing the trained model to generate sequences of pseudo-random yet interconnected instructions for hardware fuzzing. Unlike existing methods, our approach prioritizes creating interdependent data/control flow entangled instruction sequences.

\ourname, illustrated in \autoref{fig:overview}, comprises several components. The \textit{LLM-based Input Generator} generates instruction sequences for fuzzing the targeted CPU. Details about this component are discussed in \autoref{design_mld} and \autoref{design_tllm}. The \textit{RTL} and \textit{ISA Simulators} execute the given inputs on the targeted CPU and its golden model, respectively, while recording execution traces. For each test input the RTL Simulator reports coverage information, which is utilized by the LLM-based Input Generator to optimize the input generation process. The \textit{Mismatch Detector} compares execution traces to identify mismatches or potential bugs, which are manually inspected for confirmation as elaborated in \autoref{design_hwf}. 
In the following, we elucidate the fundamental components of our approach, encompassing \textit{A) the acquisition of a training dataset for instructing the LLM model}, \textit{B) the training process of the LLM model to grasp machine language intricacies}, and \textit{C) the execution of hardware fuzzing and bug detection procedures.}

\subsection{Machine Language Dataset} \label{design_mld}

A major challenge in training LLMs is the need for an extensive training dataset. While collecting data for natural languages like English is relatively easy, it becomes much more complicated for machine languages. To explore this issue, we will investigate two key questions: How do we collect a machine language dataset? And how do we represent the machine language data set for LLMs?

\subsubsection{Training Data Collection} We have two options for collecting a machine language dataset. 
i) Dynamic data collection. Recording instructions as a program runs is convenient; however, it faces challenges that disrupt data inter-dependency, such as context switches and kernel-related instructions. Rarely executed code sections may also be missing due to conditional constraints, affecting data completeness and interdependence. These issues are more pronounced when collecting data from complex programs, e.g., the Linux kernel, where repetitive instructions and infrequent execution of critical sections add complexity to data collection and interdependence.
ii) Static data collection directly gathers training data from fixed sources like GUI-compiled code, avoiding dynamic program execution complexities such as context switches and kernel-related instructions. This approach keeps the collected data isolated from OS concurrent tasks' interference, preserving intrinsic data relationships as coded in the source.
Static data collection also comprehensively captures all code segments, including rarely executed blocks, without relying on their activation during program runtime. 

In this work, we opted for static data collection as it effectively overcomes the challenges posed by instruction inter-dependency and code block rarity encountered in dynamic data collection.

\subsubsection{Training Data Representation}
This step is challenging due to several factors, including the presence of metadata (such as headers and linking information) within machine codes resulting from program compilation. Metadata can introduce complexity and ambiguity, potentially hindering the LLM model's ability to learn the language effectively and maintain the meaningfulness and interdependence of the training dataset.
To address this challenge, as illustrated in \autoref{design_tllm}, we disassemble the binaries generated from program compilation and automatically identify the start and end locations of functions within the disassembled files. We then include the machine code of each function as an individual entry in the training dataset designed for our LLM model. This approach ensures that each function, being associated with a distinct responsibility and meaning, contributes to the creation of a training set characterized by a high degree of inter-dependency among the instructions and their sequencing.

\subsection{Training of LLM Model} \label{design_tllm}
\autoref{fig:system} depicts the ML subsystem. Our approach involves a structured three-step pipeline, each phase dedicated to training the language model, advancing steadily towards our ultimate objective.

\subsubsection{Initial Training} 
In this step, the model is initialized and trained using the collected dataset. This step aims to learn the language utilized by the CPU.
For this purpose, we train a tokenizer on the full ISA. The tokenizer then prepares the inputs for the model as shown in \autoref{fig:system}(1).  

\subsubsection{Model Language Cleanup} 
Once the initial training is completed, the model can commit numerous errors in the text generation (e.g., wrong/illegal combinations of instructions). Therefore, a refinement phase is crucial. Hence, at this stage of the pipeline, \autoref{fig:system}(2), our goal is to clean up the generations of the trained model, enforcing the correct instruction associations to minimize the number of ineffective generations. 
For this purpose, we designed an RL process that leverages, as a reward agent, the ISA disassembler (cf., \autoref{design_mld}).
We avoid using, as commonly done, a probabilistic scorer, such as a neural network, for the rewarding task to prevent uncertainty and reduce errors. Employing a deterministic reward agent, we can provide the model with more precise guidance during the training, leading to better optimization policies and more precise model updates. 
This step helps avoid unnecessary CPU simulation of bad/malformed data and thus improves the overall performance of our fuzzer.

\subsubsection{Model Optimization} 
Finally, we aim to improve the training of our LLM to achieve our goal, which is generating sequences of pseudo-random yet interconnected instructions that lead to better CPU coverage. To do so, we employed another RL-based training step, utilizing a deterministic reward agent similar to the previous step. In this case, the reward function embeds the scores provided as fuzzing loop feedback comprising hardware coverage information collected during the simulation of the generated data on the targeted CPU as shown in \autoref{fig:system}(3).

We performed the previous steps for RISC-V ISA. However, it is worth noting that the approach described above is generalizable to any CPU architecture. 

\subsection{Hardware Fuzzing and Bug Detection} \label{design_hwf}
After training the LLM model (cf., \autoref{design_tllm}), we initiate the fuzzing loop.
As delineated in \autoref{fig:overview}, the LLM model generates a batch of test inputs, where each entry represents a list of instructions. These entries are then executed on the golden model and the targeted CPU using the ISA and RTL Simulators, respectively. The resulting two execution traces of each entry are analyzed by the Mismatch Detector to identify traces' discrepancies, which are documented for subsequent manual inspection as part of the bug detection process.

Additionally, the RTL Simulator reports hardware coverage metrics to the Coverage Calculator, which computes standalone, overall, and incremental coverage values for each entry as described in \autoref{imp}. These values are then used to score each entry generated by the LLM model, leading to a precise evaluation of the entries and guiding the LLM model to generate further inputs that have the potential to enhance the coverage.

% =======================================
% =======================================
% =======================================

\section{Implementation} \label{imp}
In this section, we will provide details on the implementation of \ourname components.
We deployed Synopsys VCS and the Spike simulator for RTL and RISC-V ISA, i.e., the golden model, simulations. Additionally, we developed custom components for Mismatch Detection and Coverage Calculation. 

\subsection{Mismatch Detection}
This component uses differential testing to flag potential vulnerabilities in the targeted CPU. 
It compares the architectural state changes between the targeted CPU and its golden model when both run the same input and compiles a report with uniquely identified discrepancies.
Thus, effectively reducing the manual workload for verification engineers. This is particularly advantageous when multiple instances of the same bug generate numerous mismatches.
Further, verification engineers can add filters to the Mismatch Detector in the form of architectural state values that will allow filtering out most of the false positive mismatches and accelerate vulnerability detection. 

\subsection{Coverage Calculation}
This component is responsible for receiving the coverage reports from the RTL simulator, i.e., Synopsys VCS in our implementation. Subsequently, the coverage reports undergo parsing, facilitating the calculation of three key values: stand-alone coverage, incremental coverage, and total coverage, for each coverage metric.
\textit{Stand-alone} coverage indicates the number of coverage points attained by the input under consideration. \textit{Incremental} coverage gauges the quantity of newly achieved coverage points by the current input compared to the total coverage points recorded in the previous batch. Meanwhile, \textit{total} coverage encapsulates the cumulative tally of coverage points attained thus far, incorporating the contributions of all inputs generated by the LLM model. These values are deployed in the calculation of scores assigned to each test input generated by the LLM-based input generator, thereby facilitating a comprehensive evaluation of the generated inputs, i.e., test inputs, with respect to their coverage effectiveness.

\subsection{LLM-based Input Generation}
The ML part of \ourname was fully implemented in Python with the use of the frameworks Pytorch (\url{www.pytorch.org}) and Huggingface (\url{www.huggingface.co}). The use of Huggingface is considered the standard for NLP-related tasks. Specifically, we leveraged its implementations of the tokenizer, the large language model (more precisely, of GPT2 family), and the PPO algorithm for the RL pipeline. All the experiments were conducted on a high-performance server. In the following, we describe the main steps designed for achieving our goal, principally depicted in \autoref{fig:system}.

\subsubsection{Initial Training}
The initial step in designing an NLP pipeline is defining the dictionary and its corresponding tokenizer. The tokenizer translates words (i.e., instructions) into tokens by encoding input text into an array of dictionary word indices, always serving as an intermediary step between the dataset and the language model. Decoding, on the other hand, translates an array of tokens back into text (i.e., sequence of instructions).
Next, we trained the selected model to understand the inner workings of the machine language, including grammar and instructions relationships. 
During the training, the model receives an input fragment of valid test vectors from our collected dataset, resembling $\sim$ 500K test vectors extracted by compiling the Linux Kernel, and learns how to complete it.

\subsubsection{Model Language Cleanup}
After the initial training, the model is able to utilize the CPU's language. However, having the full ISA available as a dictionary, the model will easily commit errors, generating illegal associations of instructions that a disassembler can easily detect. To overcome this limitation, which would significantly impact the quality of the end generations, we decided to perform training through a PPO-based RL, where the scoring agent is the RISC-V disassembler. The reward function is designed in such a way that correct generations are incentivized, and generations with illegal instructions are as penalized as many invalid instructions are present in the generated test vector:
\begin{equation}
    f(\text{GenText}_i) = N_i - 5*\text{Invalid}_i\
\end{equation} 
where $N_i$ is the number of instructions generated at time $i$ for GenText$_i$, and Invalid$_i$ is the number of invalid instructions present in GenText$_i$.

For the training, we utilized a dataset of 51.2K samples extracted from the larger main dataset.  
For each sample, we randomly selected the initial 2 to 5 instructions as input for the LLM. The model then completes the test vectors using its learned logic.

The training consists of 30 epochs. We monitored the PPO algorithm's loss, the Kullback-Leibler divergence between optimization policies, and the mean rewards assigned at each step to assess the training progress.

\subsubsection{Model Optimization}
Once the model went through two steps of training and the number of errors in the generations was sensibly reduced, we proceeded with the final training, where we wanted to carefully drive the model towards the exploration of the targeted CPU (i.e., increasing the reference coverage) through a PPO-based RL process. In this case, the reward function, based on the values reported by the Coverage Calculator, takes into account the overall knowledge of architecture until the $i$-th step, the incremental coverage (i.e., whether there was an improvement), and stand-alone coverage (i.e., coverage of the $i$-th sample).
In practice, the reward function guides the search direction toward generations that increase the coverage by giving a bonus and penalizing (i.e., assigning a negative reward) those that do not produce any improvement. This reward function, ultimately, pushes the model to explore more in the direction of interesting generations.
Moreover, analogously to the previous step, this training takes place with the same strategy. We utilize the same sampled dataset of 51.2K samples as input. In this case, the training is designed to last at most 15 epochs, during which the values reported by the coverage calculator are used for the reward computation.

% =======================================
% =======================================
% =======================================

\section{Evaluation} \label{eval}
We used ten instances of Synopsys VCS as a simulator and measured the effectiveness of our solution using the condition coverage metric provided by Synopsys VCS. It is imperative that this feedback captures new hardware behavior and functionalities during fuzzing. Condition coverage aligns with this goal, correlating the satisfaction of hardware design conditions with realizing new functional behaviors. An exemplary instance is fulfilling conditions leading to privilege-level transitions, such as shifting from the user to the supervisory level.
We have chosen the widely utilized RISC-V RocketCore and Boom processors, renowned as preeminent open-source processors within the RISC-V ecosystem.
In evaluating RISC-V processors, we employed the Chipyard simulation environment, which facilitates the assessment of diverse processors and ensures a uniform testing arena. Each experiment was executed over 24 hours and repeated three times to underscore the robustness and consistency of our findings.

\subsection{Design Coverage}\label{coverage}
\begin{figure}[!t]
    \centering
    \includegraphics[width=0.95\linewidth]{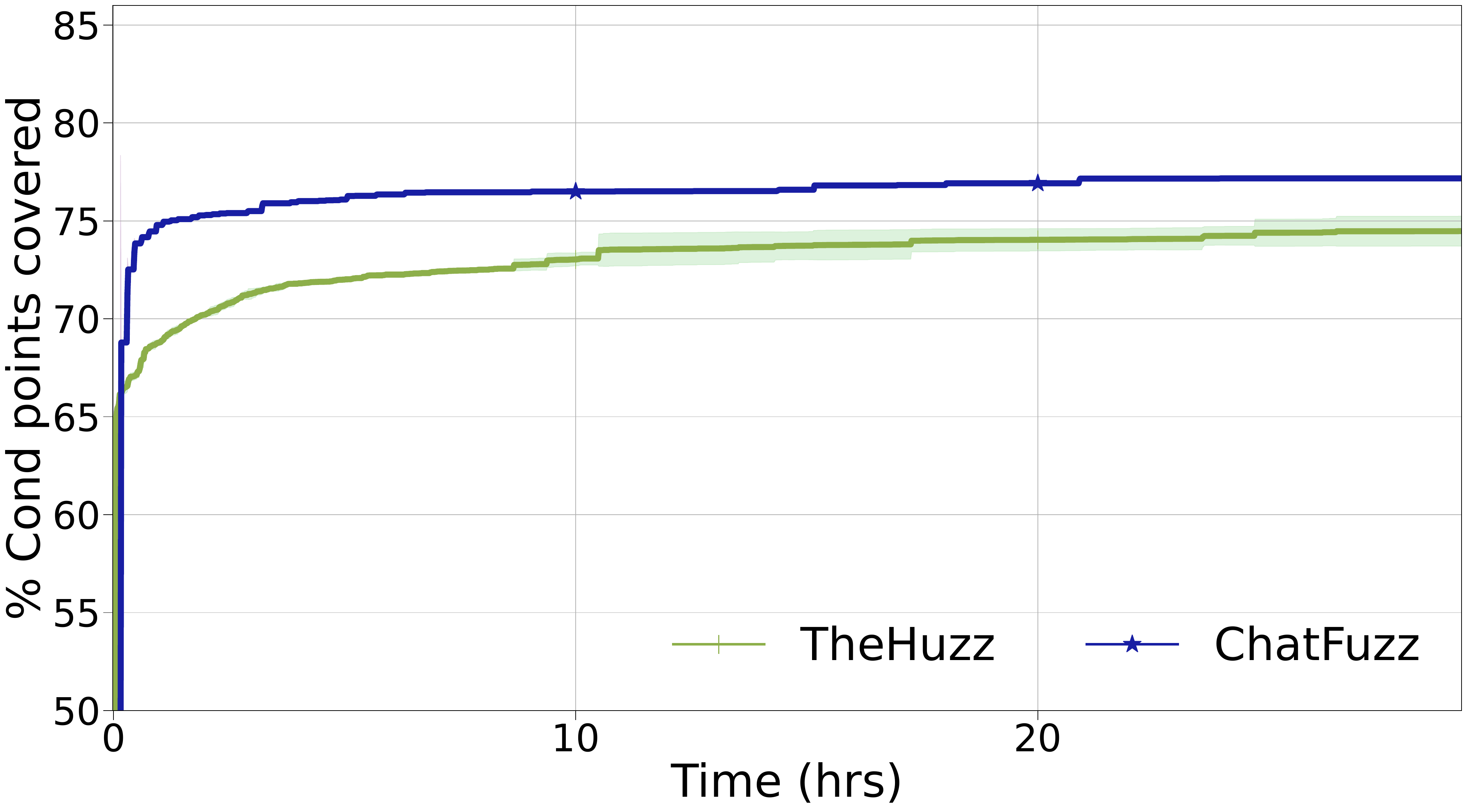}
    \caption{Coverage analysis of TheHuzz \cite{thehuzz} and \ourname over time for RocketCore.}
    \label{fig:time_cov}
\end{figure}

Our analysis revealed that both \ourname and TheHuzz incur similar runtime overhead. Nevertheless, when considering an equivalent number of generated tests (1.8K) with same number of instructions, \ourname achieved a condition coverage of 74.96\%, while TheHuzz reached 67.4\%. Remarkably, TheHuzz required around 30 hours to reach a 75\% coverage rate, i.e., \ourname achieved the same amount of coverage 34.6$\times$ faster. 
Ultimately, \ourname achieved a condition coverage rate of 79.14\% by generating 199k test cases, while TheHuzz \cite{thehuzz} attained a condition coverage rate of 76.7\% for the same number of test cases.
Furthermore, \ourname accomplishes a remarkable 97.02\% condition coverage in 49 minutes while running experiments on the Boom processor. \autoref{fig:time_cov} provides visual representation of the condition coverage for \ourname and TheHuzz during 24 hours of RocketCore fuzzing.

\subsection{Findings}
In the initial stage of our mismatch detection process\ourname effectively identified 5,866 instances of disparities within the execution traces originating from the RISC-V ISA simulator and the RocketCore. Subsequently, these identified mismatches underwent a secondary filtration process,
separating more than 100 unique mismatches. This filtration process was executed in an automated fashion. Following this, we embarked on a detailed manual analysis of these unique mismatches, the summaries of which are presented below.

\subsubsection{Bug1} According to the RISC-V specification \cite{riscv_spec}, when there are modifications made to the instruction memory, it is imperative for the software to manage cache coherency through the utilization of the \texttt{FENCE.I} instruction. Neglecting this cache coherency management can lead to unforeseeable consequences, wherein processors may rely on outdated data and execute instructions incorrectly. During testing with a generated input program by our fuzzing tool that modified the instruction memory but did not incorporate the \texttt{FENCE.I} instruction, an inconsistency was identified in the trace logs of the RocketCore processor and Spike. This disparity could have been prevented if the RISC-V specification or the RocketCore processor could detect violations of cache coherency at the hardware level. This bug has the potential to introduce cache coherency problems in software executed on the RocketCore processor, which might go unnoticed if the \texttt{FENCE.I} instruction is misused, ultimately resulting in a memory and storage vulnerability identified as CWE-1202.

\subsubsection{Bug2} 
RISC-V specification consists of arithmetic instructions such as multiply and divide~\cite{riscv_spec} that compute a value using the operand registers and update the result in the destination register.
The RocketCore processor and ISA simulator behave accordingly when executing the multiply and divide instructions.
However, the tracer module in RocketCore is not outputting the write to the destination register in RocketCore's trace output, resulting in Bug2. 
This bug may not have security consequences as it is present in the debug components of RocketCore. 
However, bugs like this can mask other security vulnerabilities that can otherwise be detected with the correct trace output information (CWE-440).

\subsubsection{Other Findings}
In conjunction with its capacity for vulnerability detection, our tool has brought to light compelling disparities between the target processor and Spike. While these disparities do not signify security vulnerabilities, they highlight the tool's capabilities in comprehensively examining the target processor. These discrepancies represent exceptional cases within the RISC-V specification and highlight the effectiveness of our approach in exploring the DUT search space. This is achieved by generating interdependent and data/control flow entangled instructions, as opposed to the conventional use of random instructions employed by state-of-the-art hardware fuzzers. We will elucidate the three most significant ones below.

\textbf{Finding1.} In line with the RISC-V specification \cite{riscv_spec}, when an instruction triggers multiple synchronous exceptions, the higher-priority exception is logged in the \texttt{mcause} register. The priority hierarchy established in the RISC-V privilege specification places the \texttt{Load/store/AMO address misaligned} exception above the \texttt{Load/store/AMO access fault} exception.
In our fuzz testing using \ourname, two test cases emerged. In the first, both \textit{Load access fault} and \textit{Load address misaligned} exceptions were simultaneously raised. In contrast, the second test case triggered both \textit{Store access fault} and \textit{Store address misaligned} exceptions concurrently. Notably, Spike responded with the \texttt{Load/Store address misaligned} exception, while RocketCore issued the \texttt{Load/Store address fault} exception.

\textbf{Finding2.} In another example, \ourname generated a pair of atomic instructions, such as \texttt{AMOOR.D}, in which it employed \texttt{R0} as a temporary location for loading data from memory, designated as \texttt{rd}. Interestingly, our tool observed that this atomic instruction appeared to function as expected, with \texttt{R0} receiving data—a behavior seemingly at odds with the RISC-V specification \cite{riscv_spec}.
Upon further investigation, we realized that this behavior represents a corner case within the RISC-V specification \cite{riscv_spec}. It is conceivable that developers, in pursuit of optimization, could implement the \texttt{AMOOR.D} operation within the memory controller. Consequently, if a user specifies \texttt{R0} as the destination register (\texttt{rd}) for this instruction, the memory controller may perform the atomic operation as intended.

\textbf{Finding3.} Another notable scenario relates to the behavior of the RocketCore processor, particularly in its treatment of the \texttt{R0} register, compared to the Spike ISA simulator. According to the RISC-V ISA specifications, the \texttt{R0} register is expected to maintain a constant value of zero, implying immunity to write operations. However, our analysis unveiled that in the execution traces generated by the RocketCore, there are occurrences of attempted writes to the \texttt{R0} register within specific sequences of instructions. It is important to note that this discrepancy is solely observed in the output traces and does not affect the functionality of RocketCore.

% =======================================
% =======================================
% =======================================

\section{Conclusion}
We introduced \ourname, a novel hardware fuzzer that utilizes large language models to learn machine language and generate complex, interdependent, data/control flow entangled and pseudo-random test cases. Our approach significantly improves condition coverage, reaching 74.96\% in less than an hour, compared to the 30 hours required by leading hardware fuzzers, i.e., \ourname achieved the same amount of coverage 34.6$\times$ faster. Also, in the case of Boom, \ourname accomplishes a remarkable 97.02\% condition coverage in 49 minutes. \ourname has successfully identified more than 100 unique mismatches, revealed two novel bugs, and exposed deviations in RocketCore behavior compared to the golden model, even in intricate corner cases specified in the RISC-V ISA specification. These results highlight \ourname's effectiveness in exploring processor vulnerabilities, offering a faster and more comprehensive approach to hardware security and testing.

\section*{Acknowledgement}
Our research work was partially funded by the Intel's Scalable Assurance Program, Deutsche Forschungsgemeinschaft (DFG) – SFB 1119 – 236615297, the European Union under Horizon Europe Programme – Grant Agreement 101070537 – CrossCon, the European Research Council under the ERC Programme - Grant 101055025 - HYDRANOS, and the US Office of Naval Research (ONR Award \#N00014-18-1-2058). 
This work does not in any way constitute an Intel endorsement of a product or supplier. Any opinions, findings, conclusions, or recommendations expressed herein are those of the authors and do not necessarily reflect those of Intel, the European Union, the European Research Council, or the US Government.

\bibliographystyle{plain}
\bibliography{bibshort.bib}

\end{document}